\documentclass[twoside,twocolumn,9pt]{article}
\usepackage{extsizes}
\usepackage[super,sort&compress,comma]{natbib} 
\usepackage[version=3]{mhchem}
\usepackage[left=1.5cm, right=1.5cm, top=1.785cm, bottom=2.0cm]{geometry}
\usepackage{balance}
\usepackage{widetext}
\usepackage{times,mathptmx}
\usepackage{sectsty}
\usepackage{graphicx} 
\usepackage{lastpage}
\usepackage[format=plain,justification=raggedright,singlelinecheck=false,font={stretch=1.125,small,sf},labelfont=bf,labelsep=space]{caption}
\usepackage{float}
\usepackage{fancyhdr}
\usepackage{fnpos}
\usepackage[english]{babel}
\usepackage{array}
\usepackage{droidsans}
\usepackage{charter}
\usepackage[T1]{fontenc}
\usepackage[usenames,dvipsnames]{xcolor}
\usepackage{setspace}
\usepackage[compact]{titlesec}
\usepackage{subcaption}
\usepackage[para]{threeparttable}

\usepackage{epstopdf}%This line makes .eps figures into .pdf - please comment out if not required.

\definecolor{cream}{RGB}{222,217,201}

\begin{document}

\pagestyle{fancy}
\thispagestyle{plain}
\fancypagestyle{plain}{

%%%HEADER%%%
%\fancyhead[C]{\includegraphics[width=18.5cm]{head_foot/header_bar}}
%\fancyhead[L]{\hspace{0cm}\vspace{1.5cm}\includegraphics[height=30pt]{head_foot/journal_name}}
%\fancyhead[R]{\hspace{0cm}\vspace{1.7cm}\includegraphics[height=55pt]{head_foot/RSC_LOGO_CMYK}}
%\renewcommand{\headrulewidth}{0pt}
}
%%%END OF HEADER%%%

%%%PAGE SETUP - Please do not change any commands within this section%%%
\makeFNbottom
\makeatletter
\renewcommand\LARGE{\@setfontsize\LARGE{15pt}{17}}
\renewcommand\Large{\@setfontsize\Large{12pt}{14}}
\renewcommand\large{\@setfontsize\large{10pt}{12}}
\renewcommand\footnotesize{\@setfontsize\footnotesize{7pt}{10}}
\makeatother

\renewcommand{\thefootnote}{\fnsymbol{footnote}}
\renewcommand\footnoterule{\vspace*{1pt}% 
\color{cream}\hrule width 3.5in height 0.4pt \color{black}\vspace*{5pt}} 
\setcounter{secnumdepth}{5}

\makeatletter 
\renewcommand\@biblabel[1]{#1}            
\renewcommand\@makefntext[1]% 
{\noindent\makebox[0pt][r]{\@thefnmark\,}#1}
\makeatother 
\renewcommand{\figurename}{\small{Fig.}~}
\sectionfont{\sffamily\Large}
\subsectionfont{\normalsize}
\subsubsectionfont{\bf}
\setstretch{1.125} %In particular, please do not alter this line.
\setlength{\skip\footins}{0.8cm}
\setlength{\footnotesep}{0.25cm}
\setlength{\jot}{10pt}
\titlespacing*{\section}{0pt}{4pt}{4pt}
\titlespacing*{\subsection}{0pt}{15pt}{1pt}
%%%END OF PAGE SETUP%%%

%%%FOOTER%%%
\fancyfoot{}
%\fancyfoot[LO,RE]{\vspace{-7.1pt}\includegraphics[height=9pt]{head_foot/LF}}
%\fancyfoot[CO]{\vspace{-7.1pt}\hspace{13.2cm}\includegraphics{head_foot/RF}}
%\fancyfoot[CE]{\vspace{-7.2pt}\hspace{-14.2cm}\includegraphics{head_foot/RF}}
\fancyfoot[RO]{\footnotesize{\sffamily{1--\pageref{LastPage} ~\textbar  \hspace{2pt}\thepage}}}
\fancyfoot[LE]{\footnotesize{\sffamily{\thepage~\textbar\hspace{3.45cm} 1--\pageref{LastPage}}}}
\fancyhead{}
\renewcommand{\headrulewidth}{0pt} 
\renewcommand{\footrulewidth}{0pt}
\setlength{\arrayrulewidth}{1pt}
\setlength{\columnsep}{6.5mm}
\setlength\bibsep{1pt}
%%%END OF FOOTER%%%

%%%FIGURE SETUP - please do not change any commands within this section%%%
\makeatletter 
\newlength{\figrulesep} 
\setlength{\figrulesep}{0.5\textfloatsep} 

\newcommand{\topfigrule}{\vspace*{-1pt}% 
\noindent{\color{cream}\rule[-\figrulesep]{\columnwidth}{1.5pt}} }

\newcommand{\botfigrule}{\vspace*{-2pt}% 
\noindent{\color{cream}\rule[\figrulesep]{\columnwidth}{1.5pt}} }

\newcommand{\dblfigrule}{\vspace*{-1pt}% 
\noindent{\color{cream}\rule[-\figrulesep]{\textwidth}{1.5pt}} }

\makeatother
%%%END OF FIGURE SETUP%%%

%%%TITLE, AUTHORS AND ABSTRACT%%%
\twocolumn[
  \begin{@twocolumnfalse}
\vspace{3cm}
\sffamily
\begin{tabular}{m{4.5cm} p{13.5cm} }

& \noindent\LARGE{\textbf{Real single ion solvation free energies with quantum mechanical simulation$^\dag$}} \\%Article title goes here instead of the text "This is the title"
\vspace{0.3cm} & \vspace{0.3cm} \\

 & \noindent\large{Timothy T. Duignan,$^{\ast}$\textit{$^{a}$} Marcel D. Baer,\textit{$^{a}$} Gregory K. Schenter,\textit{$^{a}$} and Christopher J. Mundy\textit{$^{a}$}} \\%Author names go here instead of "Full name", etc.

 & \noindent\normalsize{Single ion solvation free energies are one of the most important properties of electrolyte solutions and yet there is ongoing debate about what these values are. Only the values for neutral ion pairs are known. Here, we use DFT interaction potentials with molecular dynamics simulation (DFT-MD) combined with a modified version of the quasi-chemical theory (QCT)  to calculate these energies for the lithium and fluoride ions. A method to correct for the error in the DFT functional is developed and very good agreement with the experimental value for the lithium fluoride pair is obtained.  Moreover, this method partitions the energies into physically intuitive terms such as surface potential, cavity and charging energies which are amenable to descriptions with reduced models. Our research suggests that lithium's solvation free energy is dominated by the free energetics of a charged hard sphere, whereas fluoride exhibits significant quantum mechanical behavior that cannot be simply described with a reduced model. } \\

\end{tabular}

 \end{@twocolumnfalse} \vspace{0.6cm}

  ]
%%%END OF TITLE, AUTHORS AND ABSTRACT%%%

%%%FONT SETUP - please do not change any commands within this section
\renewcommand*\rmdefault{bch}\normalfont\upshape
\rmfamily
\section*{}
\vspace{-1cm}

%%%FOOTNOTES%%%

\footnotetext{\textit{$^{a}$~Physical Science Division, Pacific Northwest National Laboratory, P.O. Box 999, Richland, Washington 99352, USA Tel: +1 509 3756940; E-mail: tim@duignan.net}}

%Please use \dag to cite the ESI in the main text of the article.
%If you article does not have ESI please remove the the \dag symbol from the title and the footnotetext below.
\footnotetext{\dag~Electronic Supplementary Information (ESI) available: [details of any supplementary information available should be included here]. See DOI: 10.1039/b000000x/}
%additional addresses can be cited as above using the lower-case letters, c, d, e... If all authors are from the same address, no letter is required
\section{Introduction}
A grand challenge in theory, simulation and modeling is to accurately predict the interaction free energies between ions and other species in water. These free energies determine the density distributions of ions in equilibrium, which in turn determine a huge range of important properties of electrolyte solution. % Batteries, electrolysis, materials synthesis, desalination, nerve conduction are just the most obvious examples.
For example, absolute pKa values,\cite{Alongi2010}  and activity/osmotic coefficients\cite{Duignan2016} can be  determined from ion-ion interaction free energies, whereas surface tensions,\cite{Duignan2014b} surface forces,\cite{Parsons2010b} colloidal/protein stability\cite{Zhang2006} and surface potentials\cite{Slavchov2013} are directly related to ion-surface interaction free energies. 

These free energies are determined by a subtle balance of contributions, but one of the most important  is the change in the ion-water interaction energy. For example, as an ion approaches another ion or an interface there is a significant energy cost associated with removing water from the ion's hydration layers. Theoretical models therefore need to be carefully tested to ensure that they are correctly reproducing these ion-water interactions. Ionic solvation free energies, the free energy change associated with transferring an ion from vacuum to water, are the most direct experimental measurement of ion-solvent interactions. This is why molecular dynamics with classical interaction potentials (classical-MD) and continuum solvent models are often parameterized or tested by comparison with measured solvation free energies. \cite{Horinek2009,Duignan2013a} These free energies are also important in their own right as they  determine the partitioning of ions between different phases. 

Due to the electro-neutrality requirement, single ion solvation free energies are one of the only examples of a solvation free energy that is not directly experimentally accessible. A number of `extra thermodynamic assumptions' have been hypothesized in order to provide a convenient estimate of the single ion solvation free energy. (See Ref.~\citenum{Hunenberger2011}  or Ref.~\citenum{Duignan2017} for a discussion of some of these approaches.) Unfortunately, none of these have proven sufficiently compelling for the community to agree on, necessitating the use of theoretical methods to resolve this question. Theory has proven inadequate at this task so far, with estimates varying  by more than 50 kJ mol$^{-1}$.    Because of the importance of these energies to  physical chemistry, their conclusive determination would be a significant achievement in its own right. Additionally, the ability to reliably and accurately compute free energies of molecules in solution is a central problem of physical chemistry. A methodology capable of doing so would have a broad range of very exciting potential applications.

Another challenge is to partition the ionic solvation free energy into separate, physically meaningful terms, such as cavity formation and electrostatic interaction energies. Coarse-grained models which reproduce these separate contributions would not suffer from problems associated with error cancellation. This partitioning is also useful as it will enable us to identify which terms show a linear response and so can potentially be treated with reduced models.
The quasi-chemical theory (QCT)\cite{Hummer1998b,Rempe2000,Asthagiri2003,Beck2006,Rogers2010} is useful for this purpose. Ref.~\citenum{Rogers2010} in particular applies QCT to perform this partitioning with the AMOEBA water model. One particularly important contribution is associated with moving the ion across the surface potential at the air-water interface. This term is of the form $q\phi$. There are several different definitions of $\phi$,  which correspond to different definitions of the single ion solvation free energy. The expressions for these quantities are provided in the supplementary information (SI). A full discussion is beyond the scope of this article but is provided in Ref.~\citenum{Duignan2017} and the references therein.

%These energies help determine a vast range of phenomena such as phase partitioning, equilibrium constants, reaction rates, activity coefficients, solubilities and nucleation rates.

%The alternative methodologies currently available to calculate solvation free energies of ions all have significant limitations. Classical molecular dynamics (CMD) relies on too many fitted parameters, shows significant model dependence and generally neglects quantum mechanical effects such as charge transfer. Continuum solvent models cannot account for specific interactions with explicit water molecules in the first few hydration layers. Quantum mechanical cluster calculations rely on the harmonic approximation and assume that solvent molecules behave as if they were in vacuum.

\begin{figure*}
        \centering
	     \begin{subfigure}[b]{.47\textwidth}
                \includegraphics[width=\textwidth]{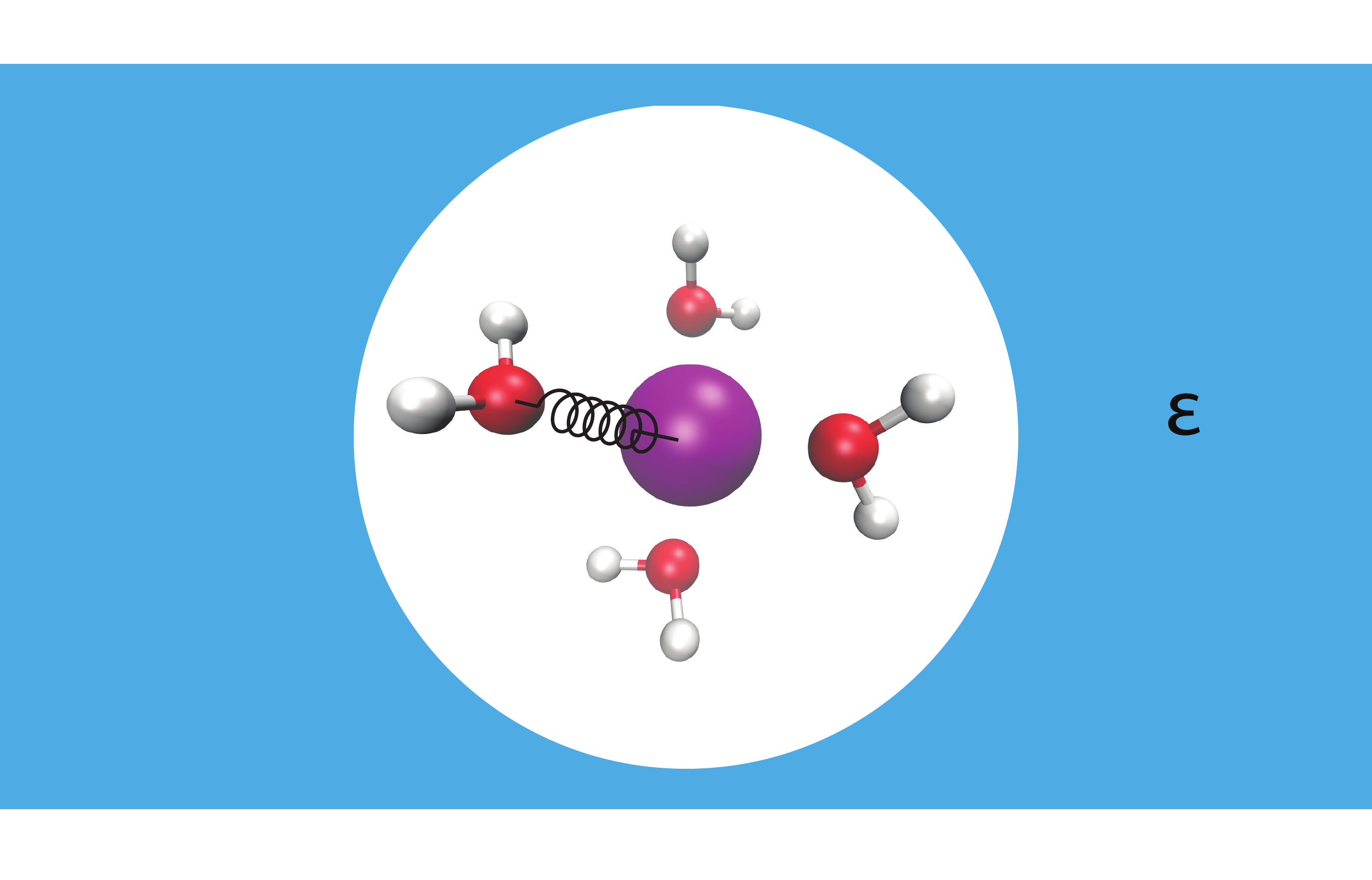}
                        \caption[]{Cluster Continuum }
        \end{subfigure}
            \begin{subfigure}[b]{.47\textwidth}
                \includegraphics[width=\textwidth]{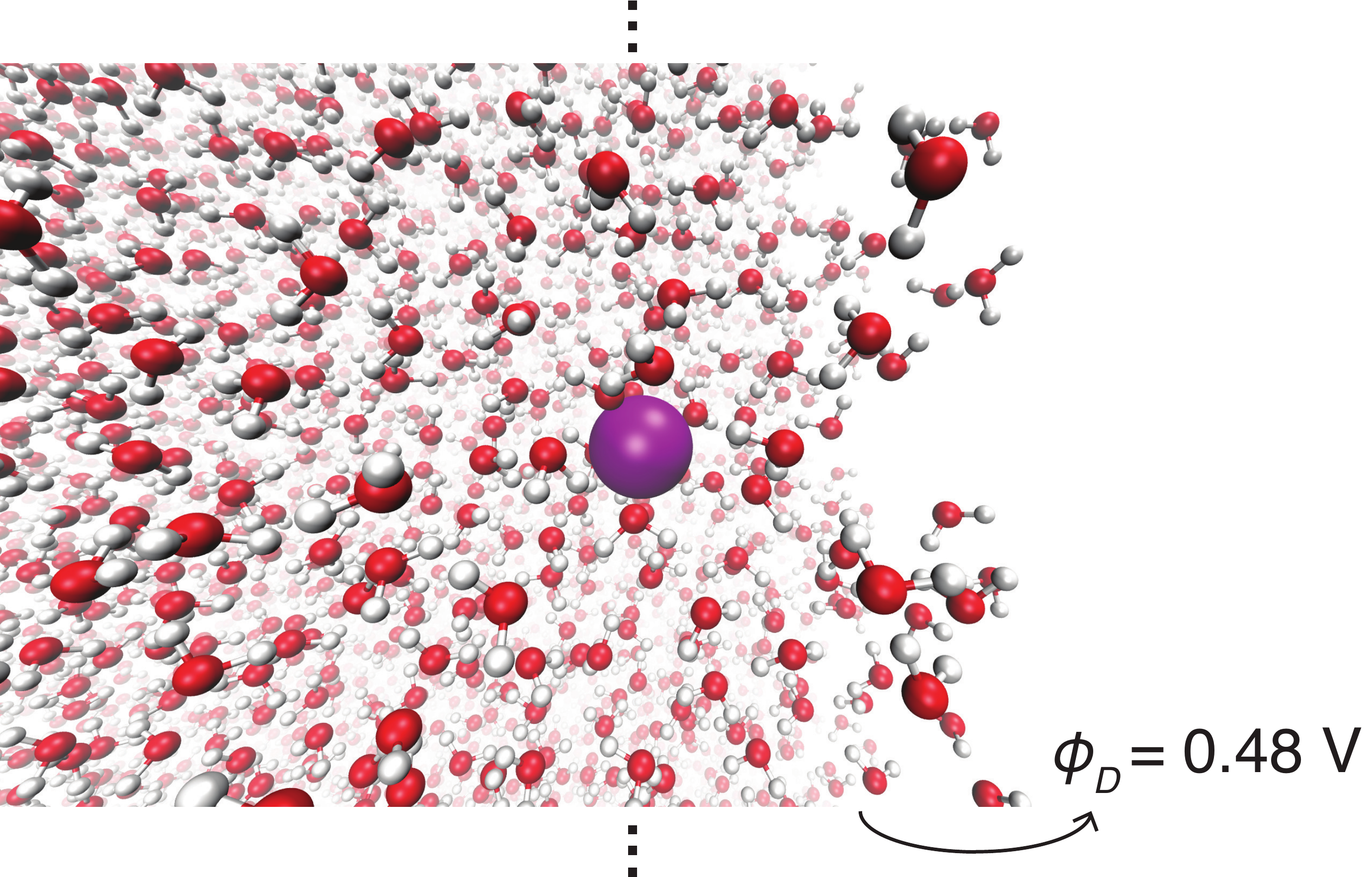}
                    \caption[]{DFT-MD}
        \end{subfigure}%
     \caption[]{Schematic depicting the two different approaches to calculating single ion solvation free energies with quantum mechanics. The cluster continuum model is the most widely used, but it relies on several  approximations and has no bulk air-water interface and so it is unclear what the surface potential contribution is. We will show how to use DFT-MD to calculate these energies including the contribution of the surface potential at the distant air-water interface.} 
         \label{approachesFIg}
\end{figure*}

%discuss cluster continuum methods?
One important approach to calculating solvation free energies of ions is the cluster continuum method. This approach combines quantum chemistry calculations on small ion-water clusters in the minimum energy geometry with a continuum solvent model.\cite{Tawa1998,Zhan2001,Asthagiri2003,Zhan2004,Tomasi2005,Bryantsev2008,Sabo2013} This approach relies on several approximations, namely anharmonicity is neglected; the contribution of the surface potential is ambiguous; and the effect of the surrounding solvent on the water structure is neglected. The validity of these approximations has has been discussed extensively elsewhere.\cite{Kathmann2007,Bryantsev2008,Rogers2011,Merchant2011,Sabo2013,Duignan2017} Figure~\ref{approachesFIg} illustrates the different approaches to calculating these quantities. 

%Another complication with this approach is that the water structures inside the dielectric cavity are essentially identical to the structures of water clusters in vacuum. The structure of water around ions in bulk liquid could well differ significantly and so this approach is not unquestionable. (CITE) In unpublished work we have attempted to generalize this methodology to larger ions, and tested the dependence on cluster size and shown that the very low error in the published values using this methodology may not be indicative of the actual error. As there is significant variation in the results indicating that this methodology is not totally reliable.

%Discuss CMD adjusted to cluster energies
Attempts at using classical-MD to address this problem have been made.\cite{Grossfield2003,Yu2010}  %Normally the parameters are adjusted to reproduce gas phase cluster energetics.
 There are significant challenges with this approach however. For example, it has recently been shown that AMOEBA  relies on  substantial cancellation of errors to reproduce ion-water dimer binding energies.\cite{Mao2016} This undermines the notion that these parameters and functional forms are transferable to the condensed phase. In addition, properties such as ionic polarizability are known to vary significantly from the gas phase to the condensed phase\cite{Serr2006} compounding this issue. As a result, problems have arisen such as the over-polarization of the chloride anion by a factor of 2 with AMOEBA compared with \emph{ab initio} calculations\cite{Rogers2010} and the unphysically large attraction of large anions to the air--water interface observed for polarizable water models.\cite{Baer2011} It remains to be seen whether a new generation of polarizable models can overcome these problems.\cite{Mao2016,Arismendi-Arrieta2016,Riera2016}
A number of recent studies have determined that density functional theory interaction potentials combined with molecular dynamics simulation (DFT-MD)  can provide an accurate description of the water structure around simple ions.\cite{Fulton2010,Baer2011a,Fulton2012,Bogatko2013,Baer2014a,Baer2016} Given the accuracy achieved in determining the local water structure around an ion, it is surprising that very few attempts to calculate solvation free energies with DFT-MD have been performed\cite{Leung2009,Weber2010} particularly given the importance of free energies in determining a range of experimentally relevant properties. It therefore remains to be seen whether this accurate structural description translates into accurate solvation and interaction free energies. In recent years significant advances in the protocols necessary to apply DFT-MD have brought us much closer to resolving this question. Herein, we establish the simulation protocols necessary to calculate single ion solvation free energies and apply these to the lithium and fluoride ions. To this end we use a modified version of QCT that goes beyond the harmonic approximation and includes the important fluctuations beyond the first hydration shell in determining accurate solvation free energies. We proceed by first calculating the solvation free energy of creating a cavity in revPBE-D3 water. We then calculate the free energy of turning a charge on in that cavity using thermodynamic integration. The free energy of replacing this charged hard sphere with a full quantum mechanically treated ion is then estimated using a free energy perturbation method, the free energy of relaxing the hard sphere repulsion and corrections associated with the use of periodic boundary conditions and a small system size are also included. Finally we estimate a correction associated with  the error in the revPBE-D3 functional. These contributions are depicted in Figure~\ref{QCTpartfig}. This allows us to arrive at real single ion solvation free energies that compare well with experiment. This methodology has the added advantage that it partitions the solvation free energies up into physically intuitive terms that can be mapped onto reduced theories for solvation. Our results suggest that lithium's solvation free energy is dominated by the free energetics of a charged hard sphere whereas fluoride exhibits behavior that requires a quantum mechanical description.

Our research highlights the importance of using DFT-MD to provide estimates for both the dipolar surface potential due to the presence of a distant air-water interface and the Bethe potential.\cite{Remsing2014} These surface potentials are essential for comparing our predictions with other published theoretical and experimental values in the literature.  Ref.~\citenum{Remsing2014} and Ref.~\citenum{Duignan2017} provide a comprehensive discussion of these surface potential. 

%Discuss experiment CPA MArcus etc.
\section{Theory}
The goal is to calculate the excess chemical potential of an ion ($X$) in water at infinite dilution:
\begin{equation}
\mu^*_{X}=-k_{\text{B}}T\ln\left< e ^{-\beta U_{XS}}\right>_{0}-E^\text{Vac}_X
\end{equation}
here we refer to this quantity as the `real' solvation free energy. Ref.~\citenum{Duignan2017} provides a detailed derivation and description of it.
 $U_{XS}$ is the ion-water interaction energy and is defined\cite{Ben-Naim1978,Ben-Amotz2005a} as $U_{XS}=U_{X,N_s}-U_{N_s}$ where  $U_{X,N_s}$ is the total energy of the solute and solvent system including the electronic energy of the ion and $U_{N_s}$ gives the total energy of a given water structure with only the water molecules present. The asterisk indicates that the `point to point' or `Ben-Naim' standard state convention is used. These values differ by -7.95 kJmol$^{-1}$ from the 1 atm to 1 M standard state often used.

Following the application of  QCT in Ref.~\citenum{Rogers2010} it is useful to partition the interaction energy of an ion in solution into a hard sphere repulsion, which creates a cavity for the ion to occupy, which is then relaxed after the ion is solvated:
\begin{equation}
U_{XS}=U_{\text{Cav}}+U_{XS}-U_{\text{Cav}}
\end{equation}
$U_{\text{Cav}}$ is a hard sphere repulsion term, which pushes only on the oxygen atoms. We use this as it allows for a simple determination of the cavity formation energy. 
%\begin{equation}
%\begin{split}
%\mu^*_{X}&=-k_{\text{B}}T\ln \left[\left<\exp^{-\beta U_{\text{\text{HS}}}}\right>_{0}\left<\exp^{-\beta U_{XS}}\right>_{U_{\text{\text{HS}}}}\left<\exp^{\beta U_{\text{\text{HS}}}}\right>_{U_{XS}+U_{\text{\text{HS}}}}\right]-E^\text{Vac}_X\\&=\mu^*_{\text{HS}}+\mu^*_{\text{QM}}+\mu^*_{\text{Relax}}
%\label{SEpart}
%\end{split}
%\end{equation}

\begin{figure}
        \centering
	     \begin{subfigure}[b]{.5\textwidth}
                \includegraphics[width=\textwidth]{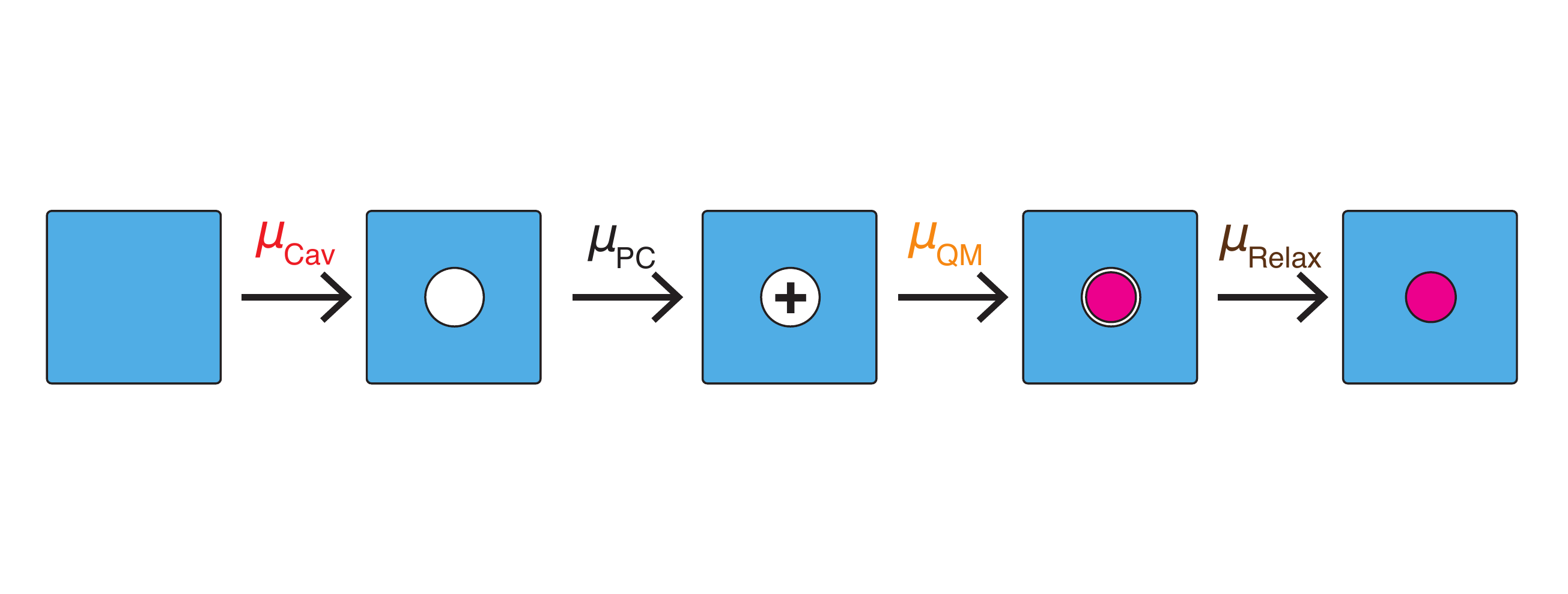}
        \end{subfigure}%
        \caption[]{Schematic of the partitioning of the single ion solvation free energies used here. The contributions are the cavity formation, point charge, quantum mechanical and hard sphere relaxation terms.}
        \label{QCTpartfig}
\end{figure}

However, instead of placing the real ion in the cavity in a one step process as is done in Ref.~\citenum{Rogers2010}, we break the process up into smaller steps. This is because in contrast to Ref.~\citenum{Rogers2010} the placement of the ion with a DFT-MD appears to be characterized by non-Gaussian fluctuations. This implies that the free energy cannot reliably be estimated using only equilibrium simulations with the ion present and not present. Instead we must break the process down into smaller steps that can be shown to be approximately Gaussian. Breaking the process up into smaller steps has the added advantage that we can identify the contributions that exhibit linear response behavior  as was done in previous studies.\cite{Remsing2016,Duignan2017} 
For these reasons we add an additional term to the partitioning which amounts to placing a point charge in the center of the hard cavity that is gradually turned on and then swapped out for the real ion. Because this charging can be performed  incrementally, the steps can be made small enough so that the assumption of Gaussian fluctuations is accurate. 
\begin{equation}
U_{XS}=U_{\text{Cav}}+U_{\text{PC}}+U_{XS}-U_{\text{PC}}-U_{\text{Cav}}
\end{equation}
where $U_{\text{PC}}=U_{\text{PC},N_s}-U_{N_s}$ is the energy change on inserting a point charge into a water structure. This partitioning is depicted in Figure~\ref{QCTpartfig}

% Crucially this energy is not simply $eq\phi$ To model a point charge in CP2K  we use a hydrogen atom with no basis functions on it with its core charge scaled to the desired value. For a classical non-polarisable water model $U_{\text{PC}}$  is simply equal to  $\phi q$ where $\phi$ is the potential at the centre of a cavity created by a hard sphere repulsion. For a quantum mechanical system however, the electrons on the respond to the presence of the point charge and so the total energy needs to be used instead. If the energies are calculated using standard Ewald  simulation, as they are with almost all simulations, then a correction needs to be made to remove the energy associated with the placing an array of charges in a neutralising background charge in order to determine a value consistent with the $\phi q$ used in classical simulation. It is also necessary to correct for the finite size effects associated with the fact that there are only a small number of water molecules availbe to solvate each ion. So the simulations is effectively being performed at a non-zero concentration, which dedends on the box size.  A valuable extension would be to partition $U_{\text{PC}}$ up into $U_{\text{PC,SR}}$ and $U_{\text{PC,LR}}$ using the error function, following the example of local molecular field theory.

We can then write the free energy of solvation as: 
\begin{widetext}
\begin{equation}
\begin{split}
\mu^*_{X}&=-k_{\text{B}}T\ln \left[\left<\exp^{-\beta U_{\text{\text{Cav}}}}\right>_{0}\left<\exp^{-\beta U_{\text{PC}}}\right>_{U_{\text{\text{Cav}}}}\left<\exp^{-\beta( U_{XS}-U_{\text{PC}})}\right>_{U_{\text{\text{Cav}}}+U_{\text{\text{PC}}}}\left<\exp^{\beta U_{\text{\text{Cav}}}}\right>_{U_{XS}+U_{\text{\text{Cav}}}}\right]-E^\text{Vac}_X\\&=\mu^*_{\text{Cav}}+\mu^*_{\text{PC}}+\mu^*_{\text{QM}}+\mu^*_{\text{Relax}}
\label{SEpart}
\end{split}
\end{equation}
\end{widetext}
%Examples of the expansion of the above averages in terms of integrals over the solvent phase space are provided in the previous paper for the case of a simple charged hard sphere. This expansion makes clear the justification for this expression.  
$\mu^*_{\text{QM}}$ gives the quantum mechanical contributions to the solvation free energy, i.e., the chemical, dispersion and exchange contributions. It accounts for the difference between the real quantum mechanically treated ion and a charged hard sphere. The electronic vacuum energy ($-E^\text{Vac}_X$) is included in this QM term. 

We can estimate the cavity formation energy  directly from simulation for cavities up to 3--4 $\text{ \AA}$  by observing the probability of cavity formation at equilibrium.
\begin{equation}
\mu^*_{\text{Cav}}=-k_{\text{B}}T\ln \left<\exp^{-\beta U_{\text{\text{Cav}}}}\right>_{0}=-k_{\text{B}}T\ln p_0(R_{\text{Cav}})
\end{equation}
where $p_0(R_{\text{Cav}})$ is the probability of finding a cavity of size $R_{\text{Cav}}$ in bulk water.
 We have provided a calculation of this term in Ref.~\citenum{Galib2016}. 

The evaluation of the point charge term ($\mu^*_{\text{PC}}$) was carried out in Ref.~\citenum{Duignan2017} where an extensive discussion of the complexities associated with the correct treatment of the surface potential terms was provided. For the purposes of this paper we calculate the Ewald solvation free energies and then make the appropriate corrections to determine estimates for the intrinsic, bulk and `real' solvation free energies. The definitions of these quantities are provided in the SI and details on how to calculate them are also provided in Ref.~\citenum{Duignan2017}. 

The point charge term can be broken into three separate contributions:
\begin{equation}
\mu^*_{\text{PC}}=\mu^*_{\text{Ch}}+\mu^*_{\text{PC(0)}}+q\phi_\text{D}+q\phi_\text{C}
\end{equation}
where $\phi_\text{D}$ is the potential created by the  dipolar orientation of water molecules at a distant air-water interface; $\phi_\text{C}$ is the potential created by the orientation of the water molecules surrounding the neutral cavity; $\mu^*_{\text{PC(0)}}$ is a correction associated with the free energy of placing the neutralized hydrogen nucleus in water (discussed in the SI); and so $\mu^*_{\text{Ch}}$ is the free energy associated with the response of the water to the charging of the ion. 
To model the point charge a hydrogen nucleus with no basis functions and with a scaled charge is used. 

This quantum mechanical term ($\mu^*_\text{QM}$) is the difference in energy for a point charge in water versus a real ion in water.\cite{Rogers2010} The complex electrostatic corrections will mostly cancel as they only depend on the charge and we can therefore simply take the difference in total energy when the point charge is replaced by the real ion. This is given by:
\begin{equation}
\mu^*_{\text{QM}}=-k_{\text{B}}T\ln \left<\exp^{-\beta( U_{XS}-U_{\text{PC}})}\right>_{U_{\text{\text{Cav}}}+U_{\text{\text{PC}}}}-E^\text{Vac}_X
\end{equation}
or its inverse:
\begin{equation}
\mu^*_{\text{QM}}=k_{\text{B}}T\ln \left<\exp^{\beta( U_{XS}-U_{\text{PC}})}\right>_{U_{\text{\text{Cav}}}+U_{\text{\text{XS}}}}-E^\text{Vac}_X
\end{equation}
We can expand the averages out with a cumulant expansion up to second order by assuming Gaussian fluctuations and performing the integral analytically.
\begin{equation}
\mu^*_{\text{QM}}\approx\left< U_{XS}-U_{\text{PC}}\right>_{U_{\text{\text{Cav}}}+U_{\text{\text{PC}}}}-\frac{1}{2k_{\text{B}}T}\left< \delta\left[U_{XS}-U_{\text{PC}}\right]^2\right>_{U_{\text{\text{Cav}}}+U_{\text{\text{PC}}}}-E^\text{Vac}_X
\end{equation}
and 
\begin{equation}
\mu^*_{\text{QM}}\approx\left< U_{XS}-U_{\text{PC}}\right>_{U_{\text{\text{Cav}}}+U_{\text{\text{XS}}}}+\frac{1}{2k_{\text{B}}T}\left< \delta\left[U_{XS}-U_{\text{PC}}\right]^2\right>_{U_{\text{\text{Cav}}}+U_{\text{\text{XS}}}}-E^\text{Vac}_X
\end{equation}
where  $\left<\delta[U]^2\right>$  simply indicates the standard deviation squared.
We can use both of these expressions and take the average to get a best estimate of this term. 

There is one complication, which is that the Bethe potential of the cell (trace of the quadrupole moment) is not precisely the same with the real ion present versus the point charge present. It is therefore necessary to include a small correction associated with the change in the Bethe potential given by $q\Delta\phi_\text{B}$ when calculating the `real' solvation free energies.(See Ref.~\citenum{Duignan2017} and the supplementary information (SI) for details) We include this correction in the charging energy term.

This method seems to work well for the lithium cation without modification. This suggests that a charged hard sphere is a good model for a lithium ion. For fluoride however, this is not the case. The anion has a large diffuse electron cloud that pushes weakly on the water molecules over a larger range so a hard sphere repulsion is a very poor  model for it and so this step is a non-linear/non-Gaussian process.  In order to make the charged hard sphere similar to the real ion we use a Born--Mayer type repulsion that acts on the oxygen atoms. 
\begin{equation}
U_\text{BM}=A\exp^{-br}
\end{equation} 
Here $r$ is the ion to oxygen distance and $A$ and $b$ are ion specific parameters. The final solvation free energy should not depend on the choice of these parameters. This process can be performed by rewriting the QM term as:
\begin{widetext}
\begin{equation}
\begin{split}
\mu^*_{\text{QM}}=&-k_{\text{B}}T\ln \left[\left<\exp^{-\beta U_{\text{BM}}}\right>_{U_{\text{\text{Cav}}}+U_{\text{\text{PC}}}} \left<\exp^{-\beta( U_{XS}-U_\text{PC}-U_{\text{BM}})}\right>_{U_{\text{BM}}+U_{\text{\text{Cav}}}+U_{\text{\text{PC}}}}\right]-E^\text{Vac}_X
\\ =&k_{\text{B}}T\ln \left[\left<\exp^{\beta U_{\text{BM}}}\right>_{U_{\text{\text{Cav}}}+U_{\text{\text{PC}}}+U_{\text{BM}}} \left<\exp^{\beta( U_{XS}-U_\text{PC}-U_{\text{BM}})}\right>_{U_{XS}+U_{\text{Cav}}}\right]-E^\text{Vac}_X\end{split}
\end{equation}
\end{widetext}
We can then break the first term up into smaller increments, gradually turning on the Born-Mayer repulsion potential so that the Gaussian approximation is accurate. 
%This requires two changes to the partitioning given above the first is the addition of an extra term associated with turning on this repulsion:
%\begin{equation}
%\mu^*_{\text{BM}}=-k_{\text{B}} T\ln\left<\exp^{-\beta U_{\text{BM}}}\right>_{U_{\text{Cav}}+U_{\text{PC}}}=k_{\text{B}} T\ln\left<\exp^{\beta U_{\text{BM}}}\right>_{U_{\text{Cav}}+U_{\text{PC}}+U_{\text{BM}}}
%\end{equation}
%We can also turn on the BM potential slowly and use TI if it is highly non-linear, i.e., non--Gaussian: 
%\begin{equation}
%\mu^*_{\text{BM}}=-k_{\text{B}} T \int_0^1 d\lambda\left<U_\text{BM}\right>_{\lambda U_{\text{BM}}+U_{\text{Cav}}+U_{\text{PC}}}
%\end{equation}
%Where 
%\begin{equation}
%\left<U_\text{BM}\right>_{\lambda U_{\text{BM}}+U_{\text{Cav}}+U_{\text{PC}}}=-k_{\text{B}}T\ln\frac{\int U_{\text{BM}}\exp^{-\beta(\lambda U_{\text{BM}}+U_{\text{PC}}+ U_{\text{\text{Cav}}}+U_{\text{Tot}}^{\text{Solv}})}d^3r_S}{\int\exp^{-\beta (\lambda U_{\text{BM}}+U_{\text{PC}}+ U_{\text{\text{Cav}}}+U_{\text{Tot}}^{\text{Solv}})}d^3r_S}
%\end{equation}
The direct and inverse estimates for all of the contributions are given in the SI, showing that the Gaussian approximation is reasonable 

The last term in Eq.~\ref{SEpart} is just the energy of relaxing the hard sphere repulsion. If the hard sphere wall is put just inside the peak in the ion-oxygen radial distribution function, then this term is quite small.\cite{Rogers2010} It is necessary to write it in the  inverse form.
\begin{equation}
\begin{split}
\mu^*_{\text{Relax}}=&-k_{\text{B}}T\ln\left<\exp^{\beta U_{\text{Cav}}}\right>_{U_{XS}+U_{\text{Cav}}}=k_{\text{B}}T\ln\left<\exp^{-\beta U_{\text{Cav}}}\right>_{U_{XS}}\\&=k_{\text{B}}T\ln  x_0(R_{\text{Cav}})
\end{split}
\end{equation}
where $x_0(R_{\text{Cav}})$ is  the probability of there being no oxygen atoms inside the hard sphere radius around the ion when the sampling is performed with the real ion-water interactions.  

Finally we account for any errors associated with the DFT functional we are using. We can do this by writing the free energy at the exact level as:
\begin{equation}
\begin{split}
\mu^*_{X}&=-k_{\text{B}}T\ln \left<\exp^{-\beta\left( U_{XS}^{\text{\text{Exact}}}-U_{XS}^{\text{\text{DFT}}}\right)}\right>_{U^{\text{DFT}}_{XS}+U^{\text{Exact}}_{N_S}}\\&-k_{\text{B}}T\ln \left<\exp^{-\beta U_{XS}^{DFT}}\right>_{U^{\text{Exact}}_{N_S}}-E^\text{Vac,Exact}_X
\label{Exactcorr}
\end{split}
\end{equation}
Here we have replaced  $\left<...\right>_0$ with $\left<...\right>_{U_{N_S}}$  to indicate that the sampling is performed with the solvent-solvent interactions turned on.
Currently computational limitations mean that the simulation must be performed with DFT level interactions, which means we must replace  $U^{\text{Exact}}_{N_S}$ with $U^{\text{DFT}}_{N_S}$  in the sampling. There is substantial evidence that, although it benefits from cancellation of errors,\cite{RuizPestana2017,Marsalek2017} revPBE-D3 does a good job describing the structure of pure water, which indicates that this is a reasonable assumption.\cite{Galib2016}  The second term then becomes the solvation free energy determined with DFT plus the DFT vacuum energy.  To estimate the first term we take advantage of the same approximation and use structures extracted from the DFT simulation. Note however that the exact expression uses DFT sampling for the ion-water interaction energy.  Hence, we do not need to assume that the ion-water interaction energy is described perfectly by the DFT level of theory as any error in this energy will be corrected for assuming it has Gaussian fluctuations. 
This is an important point as the ability of revPBE-D3 to reproduce  bulk water structure has been well tested.\cite{Galib2016,RuizPestana2017,Marsalek2017} Its ability to reproduce ion-water interactions however, is much less certain. There is strong evidence that GGA functionals accurately describes the water structure around halides\cite{Fulton2010} and divalent cations,\cite{Baer2016} but around alkali cations non-trivial discrepancies between simulation and experimental x-ray scattering and spectroscopy results have been observed. \cite{Galib2017}

The total solvation free energy can therefore be written as:
\begin{equation}
\begin{split}
\mu^*_{X}&\approx\mu^{*\text{DFT}}_{X}+\mu^*_{\text{Corr}}
\label{Exactcorr2}
\end{split}
\end{equation}
where
\begin{equation}
\begin{split}
\mu^*_{\text{Corr}}=&-k_{\text{B}}T\ln \left<\exp^{-\beta\left( U_{XS}^{\text{\text{Exact}}}-U_{XS}^{\text{\text{DFT}}}\right)}\right>_{U^{\text{DFT}}_{XS}+U^{\text{DFT}}_{SS}}\\&-\left(E^\text{Vac,Exact}_X-E^\text{Vac,DFT}_X\right)
\end{split}
\end{equation}
To estimate the exact ion binding energy ($U_{XS}^{\text{\text{Exact}}}$) we use the MP2 level of theory (the details are discussed below). As we currently lack the capability to calculate the binding energy with MP2 for the full 96 water molecule system in periodic boundary conditions we  extract approximately forty ion-water clusters from the simulation and compute the difference in ion-water binding energy for both methods with  non-periodic boundary conditions. This term converges reasonably well as the cluster size increases indicating that distant water molecules only interact electrostatically.

%Estimating this term does require the assumption that the perturbation from revPBE-D3 binding energies to the MP2 energies has Gaussian fluctuations.  It is difficult to test this assumption with DFT-MD as we cannot perform a simulation at the exact level. However, the small size of the fluctuations does indicate that any non-Gaussian behavior will not be a substantial correction.

%One alternative option would be to skip the DFT calculation and perform the MP2 correction using structures extracted from a classical simulation. We do not take this approach here however, as it is necessary to assume that the water model provides an accurate description of the water-water interactions around the ion. We are not confident that any classical water model can achieve this because they are normally parameterized and tested using bulk water and so are not guaranteed to work in the dramatically different environment around an ion.  

\section{Results and Discussion}
%real ion term show distributions and estiamte with fluctuations
%Relaxation energy term show probability curves
%Give estimates of uncertainty in each step and its determination, maybe?
\subsection{Total single ion solvation free energies}
\begin{figure}
        \centering
	     \begin{subfigure}[b]{.5\textwidth}
                \includegraphics[width=\textwidth]{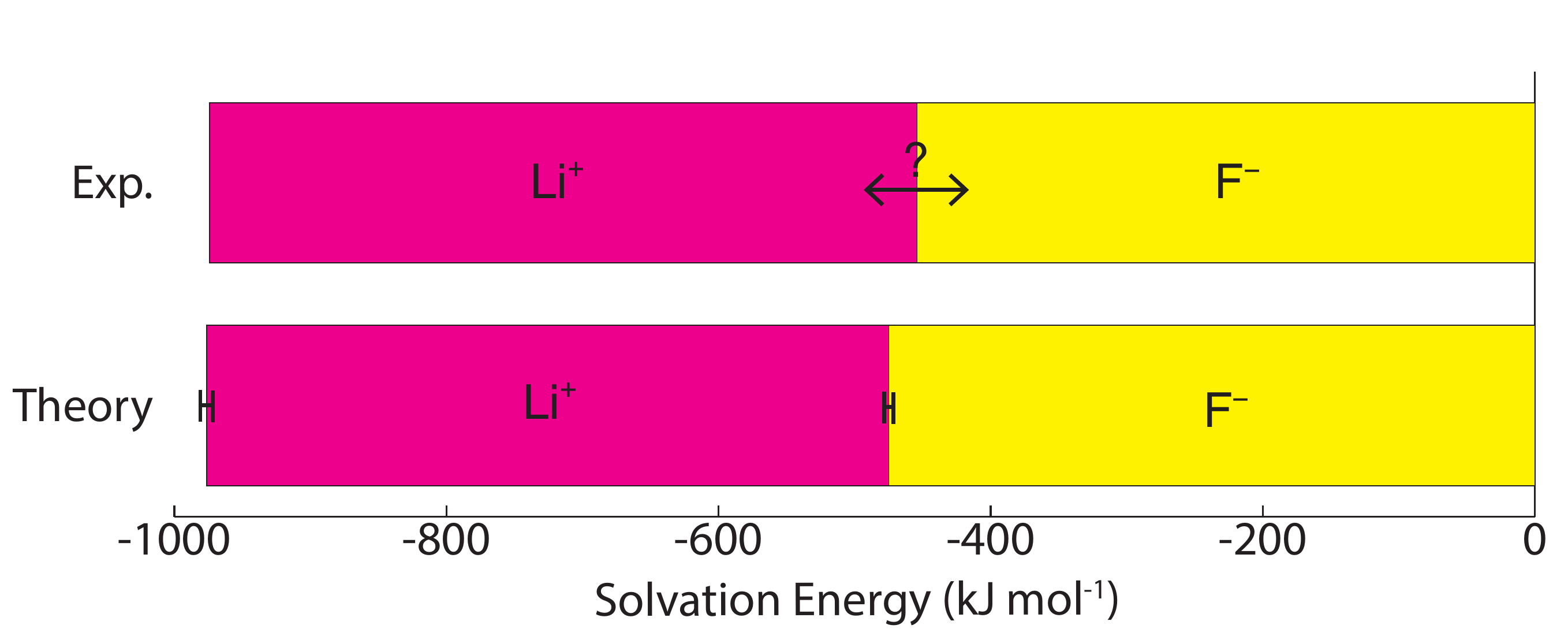}
        \end{subfigure}%
        \caption[]{Values for the `real'  solvation free energies. The spread in the experimental estimates is indicated with the double sided arrow. The statistical uncertainty in the theoretical calculation is much smaller than the spread of experimental values, highlighting why theory is useful for resolving this problem. }
        \label{SEcompfig}
\end{figure}

Table~\ref{calcSE1table} and Figure~\ref{SEcompfig} show that the final theoretical solvation free energy for the lithium fluoride pair is calculated with chemical accuracy and agrees with the experimental value within the statistical uncertainty in the calculation. 

\begin{table}
\begin{threeparttable}
\centering
\caption[]{Values for the `real'  solvation free energies. The experimental values are taken from  Ref.~\citenum{Hunenberger2011}. The division of the experimental free energy of the lithium fluoride pair into separate contributions is uncertain due to the difficulty of determining this split experimentally.  All energies are given in units of kJ mol$^{-1}$}%\begin{tabular}{lllll}
 \begin{tabular*}{0.5\textwidth}{@{\extracolsep{\fill}}llllll}
\hline
Method&Li$^+$&F$^-$ & LiF\\ \hline 
This work ($\mu^{*\text{revPBE-D3}}_{X}$) &$-498\pm3$&$-507\pm3$&$-1005\pm4$\\
This work ($\mu^{*\text{revPBE-D3}}_{X}+\mu^*_{\text{Corr}}$)&$-501\pm4$&$-475\pm3$&$-976\pm5$\\
Experiment\cite{Hunenberger2011}&$-520.1$&$-454.1$&$-974.2$\\ \hline
\end{tabular*}
\label{calcSE1table}
\begin{tablenotes}
\end{tablenotes}
\end{threeparttable}
\end{table}

These simulations were performed under bulk periodic boundary conditions using Ewald summation. Under these conditions the zero of the electrostatic potential is set so that the average potential over the cell is zero. Thus, the raw solvation free energies are not referenced to the potential in vacuum and they neglect the surface potential created by the real air-water interface.\cite{Remsing2014,Duignan2017}  We refer to these values as the Ewald solvation free energies and Table~\ref{calcSE2table} shows that these values when computed with quantum mechanics are implausible. It is well established that Ewald based free energies are not an experimentally measurable quantity due to the fact that they include a contribution from the large Bethe potential of water which has been extensively discussed.\cite{Hunenberger2011,Kathmann2011,Remsing2014,Duignan2017} 
\begin{table}
\begin{threeparttable}
\centering
\caption[]{Calculated values for the different types of solvation free energies ($\mu^*_X$) in  kJ mol$^{-1}$.}%\begin{tabular}{lllll}
 \begin{tabular*}{.5\textwidth}{@{\extracolsep{\fill}}llllll}
\hline
Ion&`Real'&Intrinsic&Bulk&Ewald\\ \hline 
Li$^+$  &$-501.4$&$-547.7$&$-519.7$&$-873.7$\\
F$^-$  &$-474.9$&$-428.6$&$-471.0$&$-91.2$\\  \hline
\end{tabular*}
\label{calcSE2table}
\begin{tablenotes}
\end{tablenotes}
\end{threeparttable}
\end{table}
 Ewald solvation free energies must be carefully corrected to account for role of the Bethe potential as well as finite cell size effects. \cite{Kastenholz2006,Duignan2017}   These corrections are used to determine the `real', intrinsic, and bulk solvation free energies, as defined in Ref.~\citenum{Duignan2017} and the SI. These values are much more in line with experimental estimates than the Ewald values.
Unfortunately, these corrections are rarely made in the context of classical-MD.  This is likely due to the fact that the Bethe potential calculated with classical-MD is normally much smaller than the quantum mechanical value.\cite{Remsing2014} ($\approx-$0.5 V compared with $\approx$ 4 V) This means that many calculations of single ion solvation free energies using classical-MD \cite{Grossfield2003,Rajamani2004,Lamoureux2006,Bardhan2012,Sedlmeier2013} are not comparable with experiments as they rely on an inherently arbitrary choice for the zero of the electrostatic potential.\cite{Duignan2017}

%\subsection{Contributions}
\begin{figure*}
        \centering
	     \begin{subfigure}[b]{.4\textwidth}
                \includegraphics[width=\textwidth]{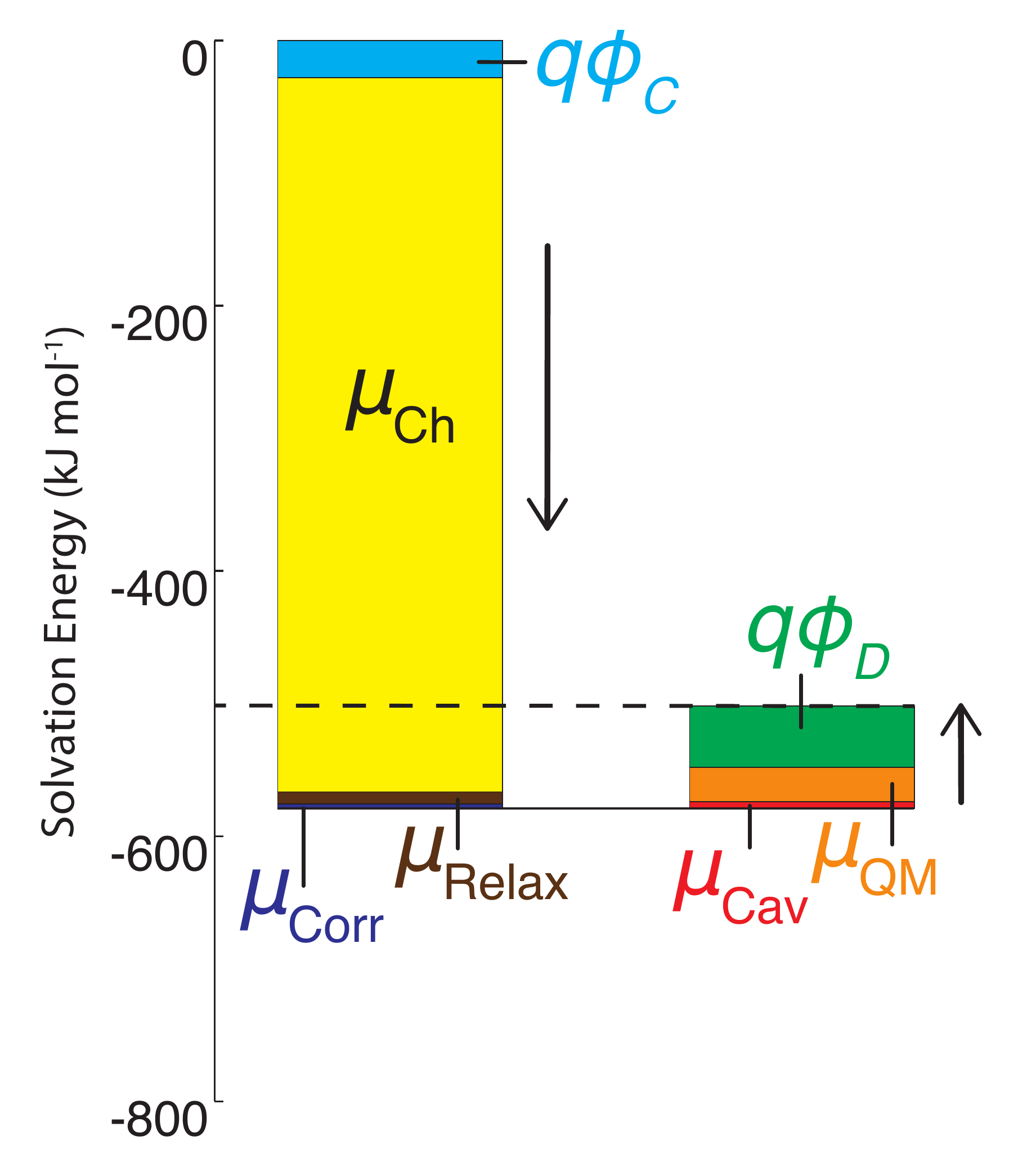}
                        \caption[]{ Lithium}
        \end{subfigure}%
            \begin{subfigure}[b]{.4\textwidth}
                \includegraphics[width=\textwidth]{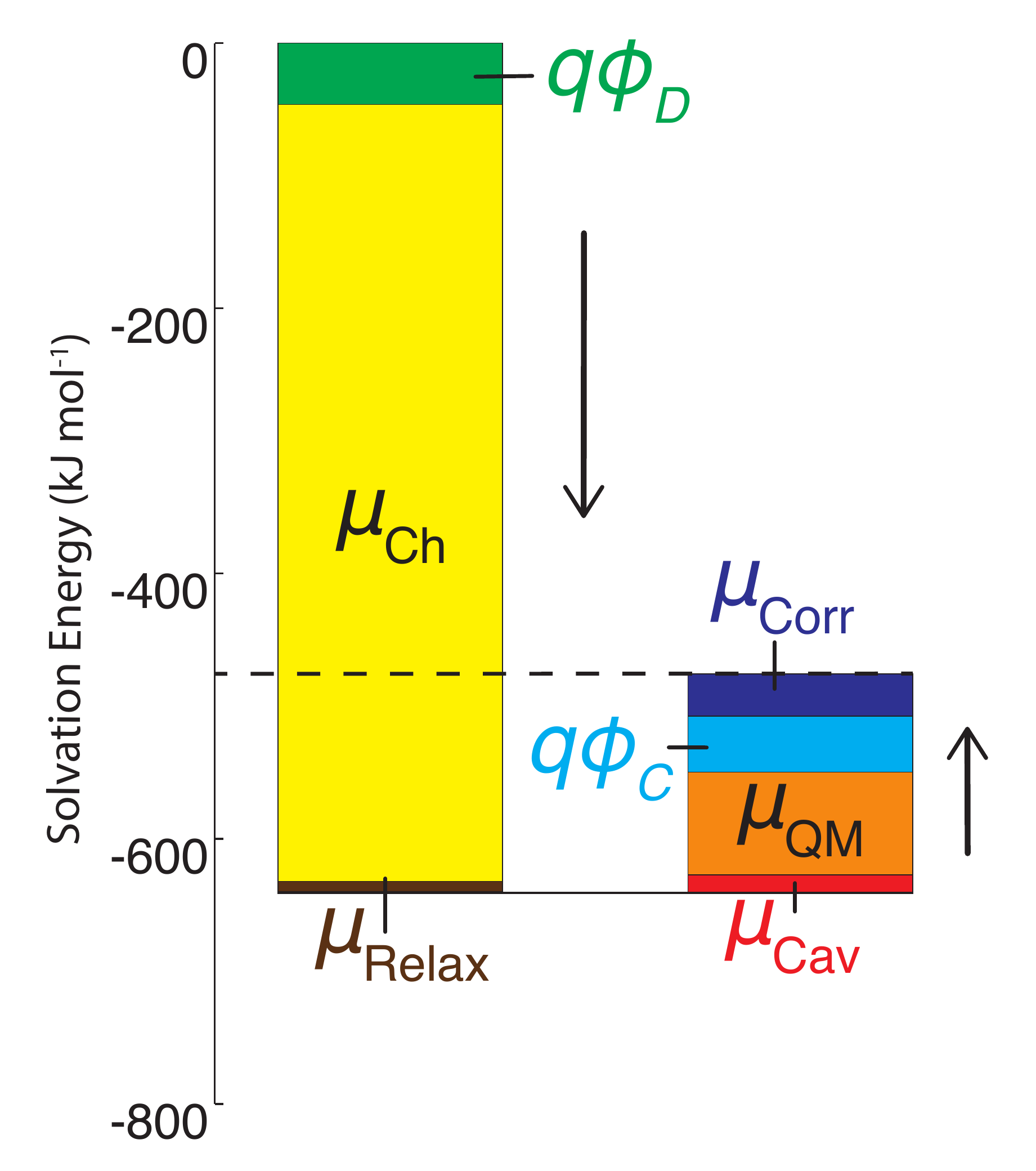}
                    \caption[]{Fluoride}
        \end{subfigure}%
     \caption[]{Contributions to the `real' solvation free energies for the fluoride and lithium ions in kJ mol$^{-1}$.}
         \label{contribsfig}
\end{figure*}

\begin{table}
\begin{threeparttable}
\centering
\caption[]{Contributions to the `real' solvation free energy ($\mu^*_X$)   for different ions in kJ mol$^{-1}$.}%\begin{tabular}{lllll}
 \begin{tabular*}{.52\textwidth}{@{\extracolsep{\fill}}llllllll}
\hline
Contribution&Li$^+$ &F$^-$ \\ \hline
$\mu^*_\text{Cav}$&$5.3\pm0.2$&$13.6\pm0.2$\\
$q\phi_D$&$46.3$&$-46.3$\\
$q\phi_C$&$-28.0$&42.5\\
$\mu^*_\text{Ch}$&$-538.6\pm3$&$-585.9\pm2$\\
$\mu^*_\text{QM}$&$25.7\pm1.4$&$77.3\pm1.9$\\
$\mu^*_\text{Relax}$&$-9.0\pm1.4$&$-7.9\pm0.7$\\
$\mu^*_\text{Corr}$&$-3.1\pm1.5$&$31.8\pm0.7$\\
\hline
\end{tabular*}
\label{contribstab}
\begin{tablenotes}
\end{tablenotes}
\end{threeparttable}
\end{table}
As stated above, the methods employed herein afford a detailed partitioning of
the solvation free energy, allowing us to connect with reduced models of
solvation.  Figure~\ref{contribsfig} and Table~\ref{contribstab} give the contributions to the single ion solvation 
free energy for lithium and fluoride.
We can see that the free energy is dominated by the charging energy, as is to be expected from a simple Born model. 
Furthermore, we have added the contributions from the surface dipole potential ($\phi_\text{D}$) and the multipolar 
cavity potential ($\phi_\text{C}$) \cite{Duignan2017} that have been discussed
in detail in previous publications.~\cite{Remsing2014,Duignan2017}
$\phi_\text{D}$ and $\phi_\text{C}$ have been demonstrated to exhibit a large
dependence on the form of molecular interaction and the corresponding local
solvation structure around the ion.  Moreover, these electrostatic potentials play a
necessary role in defining the important contributions to the solvation free
energy.  For the case of DFT-MD, $\phi_\text{C}$
and $\phi_\text{D}$ largely cancel for both ions resulting in a  small net potential.\cite{Remsing2014,Duignan2017}

An important finding of our research that can be gleaned from examining Table~\ref{contribstab} strongly suggests 
that Lithium resembles a simple charged hard sphere, 
{\emph i.e.}, lithium's $\mu^*_\text{QM}$ and $\mu^*_\text{Corr}$ terms are
quite small and can be reasonably estimated by $\mu^*_\text{Ch}$. 
%In Ref.~\citenum{Duignan2017}  we showed that a postive charged hard sphere was described reasonably accurately by a Born model 
%with the mean spherical approximation used to determine the Born
%radii.\cite{Duignan2013} This therefore implies lithium can be described approximately using a Born model. 
In contrast, fluoride has a larger charging energy than the substantially smaller lithium ion, which is then cancelled by a much 
larger $\mu^*_\text{QM}$  term. This is not unexpected, fluoride is known to have a  significantly larger exchange energy than similarly 
sized cations due to the diffuse nature of the wave function which overlaps substantially with the water molecules.\cite{Pollard2016}
This cancellation effect would be even larger if the QM term was divided into dispersion and exchange terms as these substantially 
cancel each other as well.\cite{Pollard2016} 

Although the exact partitioning of the ion solvation free energy used here has not  been applied in the case of classical-MD, 
it is possible to arrive at some general conclusions based on previous studies.
First, the cavity energy is fairly similar with both classical-MD and DFT-MD.\cite{Galib2016} The relaxation energy could be similar, assuming the classical-MD properly reproduces
the structure of water around the ion. The charge hydration asymmetry is much larger with the DFT-MD \cite{Duignan2017} 
however and in order to compensate for this classical-MD will necessarily  underestimate the charge asymmetry
in the quantum mechanical term. In particular the large exchange repulsion for the fluoride ion is almost certainly not properly
captured by the simple fitted Lennard-Jones potential often used with these models. 
 
The correction associated with the DFT functional used is non-trivial but within the expected accuracy of dispersion corrected GGAs.
The small size of the correction term for lithium is slightly misleading as it implies that the revPBE-D3 functional is very accurate 
for the lithium ion. This is not strictly the case. If we look at the correction for small lithium water clusters of the size of 
eight water molecules the correction is much larger, ($\approx-$17 kJ mol$^{-1}$) but for the larger clusters this correction is 
much smaller indicating that there is a significant cancellation of errors between the ion-water first shell and second shell 
interactions. This suggests first solvation shell water molecules are too
weakly bound whereas the more distant ones bind too strongly. 
This has been observed in other contexts, namely  DFT functionals cannot
precisely reproduce the x-ray determined peak position in 
the sodium-oxygen and potassium-oxygen radial distribution functions.\cite{Galib2017}  
Practically speaking, we observe that the cluster correction for 
lithium converges slower as a function of cluster size than for fluoride. This indicates that ion-specific interactions with the second hydration layer can be important and difficult to reproduce. This highlights a potential problem with classical-MD simulations that are often fitted to reproduce only small ion-water cluster energies and only consider first hydration layer water molecules. \cite{Grossfield2003,Arismendi-Arrieta2016,Riera2016}
The importance of second hydration layer effects has already been established on the basis of  cluster-continuum calculations.\cite{Zhan2004,Bryantsev2008}
These results also highlight the importance of accurately  modeling the full condensed
phase environment and its fluctuations in order to obtain good estimates of
solvation free energies.   

%This means that the cluster correction for 
%lithium converges slower as a function of cluster size for lithium than for fluoride, even though for fluoride the total size of 
%the correction is larger. The fact that the cluster correction needs 32 water molecules to converge highlights a problem with CMD
%methods which are often fitted to reproduce small ion-water cluster energies. 
  
To aid comparison with other studies we can use the experimentally well accepted 
difference in the solvation free energy between 
the lithium and hydrogen ions, given in 
Ref.~\citenum{Hunenberger2011}, in order to arrive at an estimate of the proton
solvation free energy. The free energy of the proton is often used as
a standard to compare different approaches and models of
solvation free energies. Table~\ref{litSEtable} provides values for the proton solvation free energy using the definitions of the `real', intrinsic and bulk solvation free energies provided in Ref.~\citenum{Duignan2017} and the SI.

\begin{table*}
\begin{threeparttable}
\centering
\caption[]{Estimates of the proton solvation free energy
  ($\mu^*(\text{H}^+$)) in kJ mol$^{-1}$. A few relevant examples from the literature are also provided for comparison. See Table 5.15 and 5.19 of Ref.~\citenum{Hunenberger2011} for a more complete list. Note that the `point to point' or `Ben-Naim' standard state convention is used.}%\begin{tabular}{lllll}
 \begin{tabular*}{1\textwidth}{@{\extracolsep{\fill}}llllll}
\hline
Source&Type&Method& $\mu^*({\text{H}^+})$ \\ \hline 
This work$^a$ & `Real'& DFT-MD &-1075$\pm3^a$\\
This work$^a$ & Intrinsic& DFT-MD &-1122$\pm3^a$ \\
This work$^a$ & Intrinsic-2$^b$& DFT-MD &-1108$\pm3^a$ \\
This work$^a$ & Bulk& DFT-MD &-1086$\pm8^a$ \\
%This work$^a$ & Ewald& DFT-MD &-1408$\pm6^a$ \\
\citeauthor{Hunenberger2011}\cite{Hunenberger2011}&`Real'& Lit. Av. &-1095.0\\
\citeauthor{Hunenberger2011}\cite{Hunenberger2011}&Intrinsic& Lit. Av.& -1108.0\\
\citeauthor{tissandier1998}\cite{tissandier1998,camaioni2005,Kelly2006}&--$^c$&Cluster Exp. (CPA)&-1112.5 \\
\citeauthor{Marcus1985}\cite{Marcus1985} &Bulk&TATB& -1064.0\\
%\citeauthor{Bryantsev2008}\cite{Bryantsev2008}&-&Cluster Theory&-1115.9 \\
\citeauthor{Zhan2001}\cite{Zhan2001}&--$^c$&Cluster Theory$^d$&-1105.8\\
\citeauthor{Asthagiri2003} \cite{Asthagiri2003} &--$^c$&Cluster Theory (QCT)$^e$&-1065.2\\
\citeauthor{Pollard2014a}\cite{Pollard2014a}&`Real'&Mix&-1105.4\\ 
\citeauthor{Pollard2014a}\cite{Pollard2014a}&Bulk&Mix&-1066.8\\ 
%\citeauthor{Leung2009}\cite{Leung2009}&Intrinsic&DFT-MD&-1108$^b$\\ 
%\citeauthor{Lamoureux2006} \cite{Lamoureux2006} &`Real'&SWM4-DP (pol-CMD)&-1093.5\\
%\citeauthor{Lamoureux2006}\cite{Lamoureux2006} &Ewald &SWM4-DP (pol-CMD)&-1041.4\\
%\citeauthor{Grossfield2003}\cite{Grossfield2003} &Ewald& AMOEBA (pol-CMD)&-1064.4 \\
%\citeauthor{Parfenyuk2011}\cite{Parfenyuk2011} &Real&?&-1092.7\\
%\citeauthor{Schmid2000}\cite{Schmid2000}&?&?&-1050?\\
\hline
\end{tabular*}
\label{litSEtable}
\begin{tablenotes}
$^a$ Error is estimated from statistical error in simulation. $^b$ Estimated using the center of nuclear charge as the molecular center. $^c$ It is unclear how the cluster based values map onto the definitions provided here.  $^d$ Ref.~\citenum{Bryantsev2008} provides a discussion of cluster-continuum theory methodology generally. $^e$ Ref.~\citenum{Asthagiri2010} provides a discussion of  cluster-continuum QCT calculations.

\end{tablenotes}
\end{threeparttable}
\end{table*}

This table shows our estimate of the `real' and intrinsic solvation free energy differ from the
average experimental estimates by $20$ kJ mol$^{-1}$ and $ -14$ kJ mol$^{-1}$
respectively. 
%This is not an unreasonable difference considering that literature estimates show a root-mean-square deviation of $\approx15$ kJ mol$^{-1}$ and $\approx50$ kJ mol$^{-1}$ respectively. (See Table 5.15 and Table 5.19 of Ref.~\citenum{Hunenberger2011}.) 

The correction from the intrinsic to the `real' solvation free
energy is determined by the dipolar surface potential, $\phi_\text{D}$. In this
study, we use the dipolar surface potential of 0.48 V calculated with DFT-MD.\cite{Remsing2014} 
The difference between the `real' and intrinsic values reported in this study is therefore much larger than the difference recommended by \citeauthor{Hunenberger2011}\cite{Hunenberger2011} who use $\phi_\text{D}=0.13$ V. 

This $20$ kJ mol$^{-1}$ disagreement for the `real' solvation free energies is particularly concerning as Table 5.15 of \citeauthor{Hunenberger2011}\cite{Hunenberger2011} shows that the  estimates of this quantity in the literature shows relatively small variation with a root-mean-square deviation of $\approx15$ kJ mol$^{-1}$ and a standard error of $\approx3$ kJ mol$^{-1}$. The reason for this disagreement could be associated with the calculation of the air-water surface potential. In order to correctly estimate  `real' solvation free energies it is necessary to know the dipolar surface potential of the air-water interface. The revPBE-D3 and BLYP-D2  functionals give a value of 0.48 V for this quantity\cite{Remsing2014}  if the oxygen atom is chosen to be the center of the water molecule.This value does not show any significant basis set or system size dependency, but there could be quantitative errors in this quantity associated with the use of generalized gradient approximation functionals. The MB-Pol water model\cite{Medders2014} has a dipolar surface potential of $\approx0.3$ V, using the oxygen atom as the molecular centre. This model has been shown to agree well with SFG measurements of the air-water interface.\cite{Medders2016} These measurements are very sensitive to the orientational structure of water molecules at the interface and so this provides some indication that the real dipolar surface potential is $\approx$ 0.2 V smaller than the revPBE-D3 value, which would explain this discrepancy. 

The intrinsic solvation free energies do not depend on the properties of the air-water interface and so the discrepancy of these values with Ref.\citenum{Hunenberger2011} cannot be explained by an incorrect surface potential. The  intrinsic solvation free energies do however depend on the inherently arbitrary choice of the oxygen atom as the center of the water molecule.\cite{Duignan2017} 
An alternative choice for the origin of the water molecule will  result in a different value for 
the intrinsic solvation free energy and so any comparison of this quantity with 
experiment must be treated with caution. For example, we can make a potentially more reasonable 
choice for the center of the water molecule, namely the center of nuclear charge. 
This definition increases the Bethe potential by  $0.14$  V and
reduces $\phi_\text{D}$ by 0.14 V. This alters the intrinsic solvation free energy (Intrinsic-2)  
by 14 kJmol$^{-1}$ and brings the theory into good agreement
with the intrinsic solvation free energy reported by \citeauthor{Hunenberger2011}.\cite{Hunenberger2011} 
The cluster pair approximation (CPA) is one of the more widely 
accepted approaches for determining single ion solvation 
free energies.\cite{tissandier1998}  It is desirable to know whether the CPA
estimate is reliable and what type of solvation free energy it is 
estimating. \citeauthor{Hunenberger2011}\cite{Hunenberger2011}  argue that their intrinsic solvation free energy is consistent with the CPA. The agreement of out Intrinsic-2 value with this value  therefore suggests that the solvation free energies 
determined with the CPA are equivalent to the solvation free energies assuming that the water 
molecules at a distant air-water interface are isotropically oriented about the center of nuclear 
charge, namely $\phi_\text{D}=0$. 

Additional evidence for this interpretation of the CPA can be inferred from Ref.~\citenum{Vlcek2013}. This paper calculated the free energies of forming small ion water clusters with the SPC/E water model and then combined these energies with the CPA to estimate single ion solvation free energies. These calculations showed that the surface potential that was consistent with the SPC/E based CPA estimate of the solvation free energy was 0.16 V less than the actual surface potential of SPC/E water. This difference is similar to the dipolar surface potential of the SPC/E water, which is + 0.15 V if the center of nuclear charge is used to determine the water molecules origin. 

The estimate for the proton solvation free energy provided here does not depend on any fitting 
to experiment or adjustable parameters (other then the single parameter used in the development of 
the revPBE functional). More importantly, it does not rely on the harmonic
approximation. The methods used in this study include complex 
electron correlation effects and do not require any unjustified assumptions about the structure 
of water around ions such as hydration numbers, as these are self-consistently
determined by DFT-MD. 
The most significant approximation is the use of 
revPBE-D3 for the contribution of  water-water interactions to  the structure of water around the 
ions. %This therefore constitutes the most rigorous determination of the single ion solvation free energy currently available in the literature. 

\subsection{Uncertainty and future work}
These error bars should be considered approximate as they are mainly determined from the differences between the direct and inverse forms of the PDT formula. Relying on direct thermodynamic integration combined with longer trajectories would be required for a precise determination of the error bars. The inverse and direct estimates agree quite closely for all of the terms and these error bars can be combined, assuming they are independent, to deteremine the error in the total solvation free energies. The resulting uncertainty indicates that this estimate is close to `chemical accuracy'.

This does not account for the unquantifiable error that arises from  assuming that revPBE-D3 structures are accurate. The neglect of quantum nuclear effects is an issue that should also be addressed in future. Path integral simulations with a classical water model\cite{Wilkins2015} indicate that this effect may be on the order of 4 kJ mol$^{-1}$. This is similar in size to the uncertainty in the estimates reported here. Surprisingly this correction is positive for lithium and negative for fluoride, resulting in a much smaller correction for the salt value.

Improving the estimate of the $\mu^*_\text{Corr}$ correction will be important to test the values determined here. In particular, the CCSD(T) level of theory should be combined with larger basis sets and larger cluster sizes.
Additionally, as discussed above the calculation of the surface potential of the air-water interface relies on GGA functionals \cite{Remsing2014} and it is not possible to easily determine the error associated with using DFT for this term, as it depends on the water structure. An improved level of theory could therefore lead to a different value, which would change the cation-anion split reported here. This would not change the experimental agreement of the salt values however as this term makes compensating contributions for the cation and the anion. Performing sampling at the RI-MP2 or RI-RPA sampling level is therefore an important goal.\cite{DelBen2015,DelBen2015b}

In future the method should be generalized to other ions such as water's self ions, potassium, sodium, cesium, iodide, divalent ions, tetra-phenyl arsonium and tetra-phenyl borate. An energy decomposition analysis\cite{Lao2015,Horn2016,Mao2016} should be used to partition the quantum mechanical energy into dispersion, exchange, induction etc. The MP2 correction for ion-ion and ion-surface PMFs should be estimated. There are also complexities associated with treating an electrolyte solution in the limit of infinite dilution, which are discussed in Ref.~\citenum{Duignan2017}. Finally, coarse grained models should be fitted to reproduce the contributions so that more complex systems can be modeled cheaply.

\section{Conclusion}
We have used DFT-MD to calculate the `real' solvation free energy of the lithium and fluoride ions including a correction that accounts for the error in the ion-water interaction calculated with DFT. The resulting salt values show excellent experimental agreement and the intrinsic single ion solvation free energies agree well with experimental values based on the CPA, provided that the center of nuclear charge is chosen to be the molecular origin for the water molecule.  This calculation moves beyond older approaches that rely on the harmonic approximation and only explicitly consider interactions with the first solvation layer.

This work has important implications for simple models of electrolyte solutions that we believe should be parametrized to reproduce the values calculated herein. We have shown that the lithium ion is reasonably well approximated as a charged hard sphere because the corrections associated with the quantum mechanical nature of the ion are relatively small. In contrast, the fluoride anion has a large quantum mechanical correction that compensates for the large charging contribution. By using a simple, well defined correction to the DFT-MD single ion solvation free energies based in MP2, our research suggests the exquisite sensitivity to  ion-specific interactions with water molecules in the second hydration layer that are not properly described with gradient corrected functionals such as revPBE-D3. 

%We can justify the CSM of Ref.~\citenum{Duignan2013a} on the basis of the following argument: Because SAPT calculations reproduce the same charge hydration asymmetry as bulk values, i.e., Na and F have the same SE. Then we may make the assumption that the contributions to the asymmetry of the SAPT calculations also determine the asymmetry of the real  ions in water. If we then look at these SAPT values for small clusters an intriguing result emerges. Although there is a large electrostatic charge hydration asymmetry, when this is added to the induction energy and combined with the exchange asymmetry these terms cancel out and a relatively monotonic behavior with size is observed, and  the experimentally observed charge hydration asymmetry matches quite nicely the asymmetry in the dispersion contribution.  In essence then the model of Ref.~\citenum{Duignan2013a}  takes advantage of the on this somewhat fortuitous cancellation modeling the ES+ind+Exch as a purely size dependent and symmetric Born model with a surface tension term. And relies on an accurate calculation of the dispersion interaction to accurately reproduce the charge hydration asymmetry. This approach is preferable to the normal CMD approach because firstly, the dispersion and exchange asymmetry do not actually cancel, so you need to include some asymmetry with this term as well, so you don not gain any simplicity by having the wrong ES asymmetry, secondly it cannot explain the unusual size behavior of the copper and silver cations which cannot be explained with CHA as they are positively charged.

\section{Calculation details}
The system contained 96 water molecules with the ion located at the center of a 14.3$^3$~\AA$^{3}$  supercell. Details for the charging free energy are given in Ref.\citenum{Duignan2017}.
 Born-Oppenheimer NVT simulations (at 300~K) were performed under PBC using  the \texttt{CP2K} simulation suite (http:www.cp2k.org) with the \emph{QuickStep} module for the DFT calculations.~\cite{VandeVondele2005} Shorter range double zeta basis sets optimized for the condensed phase\cite{VandeVondele2007} were used in conjunction with GTH pseudopotentials~\cite{Goedecker1996} and a 400~Ry cutoff for the auxiliary plane wave basis and a 0.5 fs time step.  A Nos\'e--Hoover thermostat was attached to every degree of freedom to ensure equilibration.~\cite{Martyna1992} The revised Perdew, Burke, and Ernzerhof (revPBE)\cite{Perdew1996,Zhang1998} functional with the D3 dispersion correction due to Grimme\cite{Grimme2004,Grimme2010}  was used.  The energies were accumulated for $\approx$ 12~ps after $\approx$ 2~ps of equilibration for each simulation. The details of the Bethe potential calculation are provided in Ref.~\citenum{Duignan2017}. 

For the hard sphere repulsion we use a potential of the form:
\begin{equation}
U_{\text{Cav}}=\sum_{O}1-\tanh\left(\left(r_{\text{XO}}-R_\text{Cav}\right)/0.05\right)
\end{equation}
For lithium $R_\text{Cav}$ was set to 2 \AA\ and for fluoride it was set to 2.6 \AA.
The cavity formation energy was calculated by rasterizing the cell for large revPBE-D3 slab calculations (details given in Ref.~\citenum{Galib2016}). The relaxation energies were determined from the cumulative radial distribution functions from the ion to the closest oxygen atom. (See SI for details) For the parameters for the Born-Mayer potential we use the value for $b$ determined in Ref.~\citenum{Duignan2013} and we choose several different $A$ values so that the inverse and direct estimates agree to within $2$ kJ mol$^{-1}$ for each step, which implies the fluctuations are Gaussian to a reasonable approximation. 
 
ORCA\cite{Neese2012} was used to calculate the cluster correction to the revPBE-D3 functional at the MP2 level of theory. For the revPBE-D3 calculation  \texttt{CP2K}  was used with the periodicity none option and a larger cell size to remove any box size dependence. Otherwise the same parameters, basis sets etc. as the simulation were used. Clusters of  24 water molecules were used in the cluster correction calculation with 3 frames per picosecond. These calculations showed good convergence with the 32 water molecule cluster being within 1 kJ mol$^{-1}$ of the 24 water molecule cluster. 
%For lithium the convergence was slightly slower and an exponential function was used to extrapolate to the infinite cluster size limit which led to a $+ 0.7$ kJ mol$^{-1}$ shift. 
For the MP2 calculation aug-cc-pVDZ basis sets\cite{Dunning1989,Kendall1992} were used for the hydrogen, oxygen and fluoride atoms. The cc-pCVDZ\cite{Woon1995} basis set was used for the lithium ion. The 1s orbitals on the fluoride and oxygen atoms were frozen for the mp2 level calculations. %The use of augmented double zeta level basis sets is supported by the good agreement with the counterpoise corrected quintuple zeta basis set calculations of the ion binding energy in the four water molecule clusters.  This is due in part to cancellation of errors but rigorous basis set extrapolation is not feasible with the larger clusters examined. 
There is some error associated with the basis set size and the MP2 level of theory that was used for the cluster calculations. To correct for this, the binding energies of ions to the nearest four water molecules were  calculated with the CCSD(T) level of theory and with MP2 using quadruple zeta basis sets and the counter poise correction. The average differences compared to the MP2 double zeta level calculations were used to estimate corrections for these two issues. These corrections were relatively small ($\approx$ 6 kJ mol$^{-1}$). Values for every term in the calculation that contributes to the solvation free energies are given in the SI.

\section{Acknowledgements}
We would like to thank David Dixon, Thomas Beck, Shawn Kathmann, Liem Dang, Philippe H$\ddot{\text{u}}$nenberger, Sotiris Xantheas, Richard Remsing and John Weeks for helpful discussions. Computing resources were generously allocated by PNNL's Institutional Computing program. 
This research also used resources of the National Energy Research Scientific Computing Center, a DOE Office of Science User Facility supported by the Office of Science of the U.S. Department of Energy under Contract No. DE-AC02-05CH11231.
TTD, GKS and CJM were supported by the U.S. Department of Energy, Office of Science, Office of Basic Energy Sciences, Division of Chemical Sciences, Geosciences, and Biosciences. MDB was supported by MS$^{3}$ (Materials Synthesis and Simulation Across Scales) Initiative, a Laboratory Directed Research and Development Program at Pacific Northwest National Laboratory (PNNL).  PNNL is a multiprogram national laboratory operated by Battelle for the U.S. Department of Energy.

%%%REFERENCES%%%
\bibliography{libraryAbrev}
\bibliographystyle{rsc} %the RSC's .bst file
\section{Supplementary Information}
\subsection{Surface potential definitions}
Tables~\ref{SPDef} and \ref{SEDef} summarize the definitions of the surface potentials used throughout this paper. See Ref.~\citenum{Duignan2017} for a full discussion of how to calculate these terms. 
\begin{table*}
\centering
\caption[]{Surface potential definitions}%\begin{tabular}{lllll}
 \begin{tabular*}{.5\textwidth}{@{\extracolsep{\fill}}clclc}
\hline
Type &Expression\\ \hline 
Dipolar&$ \phi_\text{D}$=$-\epsilon_0^{-1}\int_0^{z/2} dz P_z(z)$ \\
Bethe&$\phi_\text{B}$=$-\frac{1}{6V\epsilon_0} \sum_i q \left<r^2\right>_i $\\
Cavity&$ \phi_\text{C}$=See SI of Ref.\citenum{Remsing2014}\\
Net&$ \Phi^{\text{HW}}$=$\phi_\text{C}+\phi_\text{D}$\\
Total&$ \Delta\phi$=$\phi_\text{D}+\phi_\text{B}=-\epsilon_0^{-1}\int_{z_l}^{z_v} dz \rho(z)z$\\
\hline
\end{tabular*}
\label{SPDef}
\end{table*}

\begin{table*}
\centering
\caption[]{Four types of solvation free energies}%\begin{tabular}{lllll}
 \begin{tabular*}{.5\textwidth}{@{\extracolsep{\fill}}clclclclc}
\hline
Type&Expression \\ \hline 
Real &$\mu^*_{X}$ \\
Intrinsic &$\mu^{*\text{Int}}_{X}$  =$\mu^*_{X}-q_I\phi_\text{D}$ \\
Bulk &$\mu_{X}^{*\text{Bulk}}$ =$\mu^*_{X}-q_I\Phi^{\text{HW}}$\\
Ewald & $\mu_{X}^{*\text{Ewald}}$=$ \mu^*_{X}-q_I\phi_\text{D}-q_I \phi_\text{B} +\mu_\text{Ew-Corr}$ \\ \hline
\end{tabular*}
\label{SEDef}
\end{table*}

%\subsection{Ion-water structure}
%The following table compares the theoretical ion-oxygen peak position with experiment and gives the theorerical coordination number.
%\begin{table}
%\centering
%\caption[]{Ion-water structure}%\begin{tabular}{lllll}
 %\begin{tabular*}{1\textwidth}{@{\extracolsep{\fill}}clclclc}
% \hline
 %Ion & $r_\text{max} $ $(\text{\AA})$  (Exp.) &$r_\text{max} $ $(\text{\AA})$  (Theory)  &$N_C$  (Theory)   \\
%\hline
%Li$^+$ &2.06&1.97&4.0\\
%F$^-$& 2.68 &2.70 &5.8\\
%\hline
%\end{tabular*}
%\label{ionWatstrucTab}
%\end{table}

\subsection{Calculation details}
The free energy of placing a hydrogen nucleus with its charge scaled to zero is non-negligible when calculated with \texttt{CP2K}.
This is a numerical issue associated with the pseudo-potential of the core of the hydrogen atom and the dispersion interaction calculation. This term can be easily estimated using the following relationship:
\begin{equation}
\mu^*_{\text{PC(0)}}=-k_{\text{B}} T\ln\left<\exp^{-\beta U_{\text{PC}}(0)}\right>_{U_{\text{Cav}}}
\label{PCSE}
\end{equation}
 or its inverse:
 \begin{equation}
\mu^*_{\text{PC(0)}}=k_{\text{B}} T\ln\left<\exp^{\beta U_{\text{PC}}(0)}\right>_{U_{\text{Cav}}+U_{\text{PC}}(0)}
\label{PCSEinv}
\end{equation}
where $U_{\text{PC}}(0)$ is the energy change on placing the neutral hydrogen nucleus in the cavity. This term is  entirely unphysical and so we do not want to include it in the charging free energy. It will not contribute to the final energy as  its contribution will be cancelled out when the hydrogen atom is swapped out for the real ion in the following step. This contribution is therefore combined with the quantum mechanical free energy term in the results given below.

The Bethe potential was calculated for several situations to examine how much it varies. The results are presented in Table~\ref{BethePots}. 
\begin{table*}
\centering
\begin{threeparttable}
\caption[]{Bethe potentials ($\phi_\text{B}$).}%\begin{tabular}{lllll}
 \begin{tabular*}{.5\textwidth}{@{\extracolsep{\fill}}clclclclc}\hline
Solute&$\phi_\text{B}(V)$ \\ \hline
 $R_{\text{HS}}=2.0$ \AA,   $q=0$ &3.407\\
  $R_{\text{HS}}=2.0$ \AA,   $q=1$&3.409\\
 $R_{\text{HS}}= 2.6$ \AA,   $q=0$ &3.414\\
  $R_{\text{HS}}=2.6$ \AA,   $q=-1$ &3.402\\
Li$^+$& 3.417\\
F$^-$& 3.458\\
\hline
\end{tabular*}
\label{BethePots}
\begin{tablenotes}
\end{tablenotes}
\end{threeparttable}
\end{table*}
It is clear that there is a small difference when the real ion is present compared to when the charged hard sphere is present. This leads to a small correction of 0.8 kJ mol$^{-1}$ and -5.4 kJ mol$^{-1}$ to the values reported in Ref.~\citenum{Duignan2017} for the 2 \AA\ and 2.6 \AA\ charged hard sphere solvation free energies. See  Ref.~\citenum{Duignan2017} for details regarding the computation of the Bethe potential. 

To calculate the cavity formation energy we had to determine the position of the hard sphere repulsion. The hard sphere repulsion does not occur at precisely $R_\text{Cav}$ as this potential is not infinitely sharp. We therefore specify a range over which the solute-oxygen radial distribution function is increasing sharply and use the middle of this range in the cavity formation energy expression calculated in Ref.~\citenum{Galib2016}. The energy at either extreme of this range provides an uncertainty in this value. For lithium the range is 2.07 to 2.12 for fluoride the range is 2.69 to 2.71. The same cavity radius range is used to determine the energy of relaxing the hard sphere constraint and the uncertainty in this energy.

\subsection{Full Contributions}
Table~\ref{contribsLidettab} and Table~\ref{contribsFdettab} give a much more detailed breakdown of the contributions to the solvation free energies  for both lithium and fluoride respectively.  Adding all of the values in the final column gives the total `real' solvation free energies.~\begin{table*}
\centering
\begin{threeparttable}
\caption[]{Contributions to solvation free energies estimated using the direct and inverse formulations of the PDT for lithium in units of kJ mol$^{-1}$. }%\begin{tabular}{lllll}
 \begin{tabular*}{1\textwidth}{@{\extracolsep{\fill}}clclclclc}\hline
Contribution &Direct&Inverse&Final\\ \hline
$\mu^*_\text{Cav}$$^a$&5.08 to 5.50 &---&5.29  \\
$\mu^*_{\text{PC}(0)}$&$-24.49$& $-21.95$&$-23.22$\\
$\Delta\mu^*_\text{Ew}(0.1)$$^b$&$-42.05$&$-40.40$&-41.22\\
$\Delta\mu^*_\text{Ew}(0.2)$&$-51.32$&$-52.41$&-51.87\\
$\Delta\mu^*_\text{Ew}(0.3)$&$-63.50$&$-60.73$&-62.12\\
$\Delta\mu^*_\text{Ew}(0.4)$&$-73.76$&$-75.56$&-74.66\\
$\Delta\mu^*_\text{Ew}(0.5)$&$-86.12$&$-85.74$&-85.93\\
$\Delta\mu^*_\text{Ew}(0.6)$&$-97.32$&$-99.39$&-98.35\\
$\Delta\mu^*_\text{Ew}(0.7)$&$-108.58$&$-106.76$&-107.67\\
$\Delta\mu^*_\text{Ew}(0.8)$&$-116.30$&$-115.72$&-116.01\\
$\Delta\mu^*_\text{Ew}(0.9)$&$-125.03$&$-126.44$&-125.73\\
$\Delta\mu^*_\text{Ew}(1.0)$&$-136.90$&$-135.86$&-136.38\\
$\mu_{\text{Ew-Corr}}$$^c$&---&---&3.71\\
$q\phi_D$&46.31&---&46.31\\
$q\phi_B$&329.69&---&329.69\\
$q\phi_C$&-27.98&---&---\\
$\mu^*_\text{Ch-Ion}$$^d$& 48.62&$49.28$&48.94 \\
$\mu^*_\text{Relax}$$^a$&---& $-7.61$ to $-10.40$&$-9.01$ \\
$\mu^*_\text{Corr}(24)$& ---&---&  $-2.09$& \\
$\mu^*_\text{QZ-Corr}(4)$$^e$&--- &---&$-1.02$\\
$\mu^*_\text{CCSD(T)-Corr}(4)$$^f$&--- &---&$-0.06$\\
\hline
\end{tabular*}
\label{contribsLidettab}
\begin{tablenotes}
$^a$ Effective $ R_\text{Cav}=$ 2.07  to 2.12 \AA 

 $^b$ $\mu^*_\text{Ew}$ is calculated using Ewald summation and $\Delta\mu^*_\text{Ew}(q)=\mu^*_\text{Ew}(q)-\mu^*_\text{Ew}(q-\Delta q)$

$^c$ Described in Ref.~\citenum{Duignan2017}

$^d$ Free energy of replacing point charge with real ion

$^e$ Correction   to counterpoise corrected quadrupole zeta basis sets with four waters.

$^f$ Correction to  CCSD(T) calculated with four waters

\end{tablenotes}
\end{threeparttable}
\end{table*}

\begin{table*}
\centering
\begin{threeparttable}
\caption[]{Contributions to solvation free energies estimated using the direct and inverse formulations of the PDT for fluoride in units of kJ mol$^{-1}$. }%\begin{tabular}{lllll}
 \begin{tabular*}{1\textwidth}{@{\extracolsep{\fill}}clclclclc}\hline
Contribution &Direct&Inverse&Average\\ \hline
$\mu^*_\text{Cav}$$^a$&13.39 to 13.80 &---&13.60 \\
$\mu^*_{\text{PC}(0)}$&$-13.97$&$ -16.80$&$-15.39$\\
$\Delta\mu^*_\text{Ew}(0.05)$&$16.89$&$16.89$&$16.89$\\
$\Delta\mu^*_\text{Ew}(0.10)$&$14.74$&14.99&14.87\\
$\Delta\mu^*_\text{Ew}(0.15)$&$12.69$&$12.27$&12.48\\
$\Delta\mu^*_\text{Ew}(0.20)$&$9.73$&$10.52$&$10.13$\\
$\Delta\mu^*_\text{Ew}(0.25)$&$7.82$&$7.72$&$7.77$\\
$\Delta\mu^*_\text{Ew}(0.30)$&$4.42$&$5.94$&$5.18$\\
$\Delta\mu^*_\text{Ew}(0.35)$&$2.73$&$0.60$&$1.67$\\
$\Delta\mu^*_\text{Ew}(0.40)$&$-2.60$&$-1.59$&$-2.10$\\
$\Delta\mu^*_\text{Ew}(0.45)$&$-4.72$&$-6.07$&$-5.39$\\
$\Delta\mu^*_\text{Ew}(0.50)$&$-9.33$&$-10.16$&$-9.75$\\
$\Delta\mu^*_\text{Ew}(0.55)$&$-13.57$&$-11.82$&$-12.70$\\
$\Delta\mu^*_\text{Ew}(0.60)$&$-15.38$&$-15.88$&$-15.63$\\
$\Delta\mu^*_\text{Ew}(0.65)$&$-18.89$&$-18.71$&$-18.80$\\
$\Delta\mu^*_\text{Ew}(0.70)$&$-22.20$&$-22.80$&$-22.50$\\
$\Delta\mu^*_\text{Ew}(0.75)$&$-25.78$&$-25.29$&$-25.54$\\
$\Delta\mu^*_\text{Ew}(0.80)$&$-28.34$&$-29.19$&$-28.77$\\
$\Delta\mu^*_\text{Ew}(0.85)$&$-31.92$&$-30.98$&$-31.45$\\
$\Delta\mu^*_\text{Ew}(0.90)$&$-33.75$&$-34.87$&$-34.31$\\
$\Delta\mu^*_\text{Ew}(0.95)$&$-37.36$&$-36.87$&$-37.11$\\
$\Delta\mu^*_\text{Ew}(1.00)$&$-39.59$&$-41.21$&$-40.40$\\
$\mu_{\text{Ew-Corr}}$&---&---&$5.72$\\
$q\phi_D$&$-46.31$&---&$-46.31$\\
$q\phi_B$&$-333.65$&---&$-333.65$\\
$q\phi_C$&$42.44$&---&---\\
$\mu^*_\text{BM1}$$^b$&$24.02$ &$21.78$&$22.90$ \\
$\mu^*_\text{BM2}$$^b$&$$&$$&21.01 \\
$\mu^*_\text{BM3}$$^b$& 38.85&$39.75$&39.30 \\
$\mu^*_\text{ChBM-Ion}$& $9.43$&$9.57$&$9.50$ \\
$\mu^*_\text{Relax}$$^a$&---&$-7.11$ to $-8.61$&$-7.86$ \\
$\mu^*_\text{Corr}(24)$& ---&---&37.19 \\
$\mu^*_\text{QZ-Corr}(4)$& ---&---&$-2.64$\\
$\mu^*_\text{CCSD(T)-Corr}(4)$& ---&---&$-2.78$\\
\hline
\end{tabular*}
\label{contribsFdettab}
\begin{tablenotes}
$^a$ Effective $R_\text{Cav}=$ 2.69 to 2.71 \AA$^a$  

$^b$ Free energy of inserting Born-Mayer repulsion. $b=2.338$, $A=$237.5, 475 and 950 (a.u.)

\end{tablenotes}
\end{threeparttable}
\end{table*}

\end{document}